\documentclass[twocolumn,showpacs,preprintnumbers,amsmath,amssymb]{revtex4}

\begin{document}

\preprint{}

\title{Detection of the Neutrino Fluxes from Several Sources}

\author{D.L.~Khokhlov}
 \affiliation{Sumy State University}
 \email{khokhlov@cafe.sumy.ua}

\date{\today}

\begin{abstract}
It is considered the detection of neutrinos moving from the
opposite directions. The states of the particle of the detector
interacting with the neutrinos are connected with the
P-transformation. Hence only a half of neutrinos gives
contribution into the superposition of the neutrino
states. Taking into account the effect of the opposite neutrino
directions the total neutrino flux from several sources are in the
range 0.5--1 of that without the effect.
The neutrino flux from nuclear reactors
measured in the KamLAND experiment is
$0.611\pm 0.085 {\rm (stat)} \pm 0.041 {\rm (syst)} $ of the
expected flux. Calculations for the conditions of the KamLAND
experiment yield the neutrino flux taking into
account the effect of the opposite neutrino directions, 0.555,
of that without the effect that may account
for the neutrino flux observed in the KamLAND experiment.
\end{abstract}

\maketitle

Let two equal neutrino fluxes move towards the neutrino detector
from the opposite directions. The first neutrino flux moves from
the $x$ direction, and the second neutrino flux moves from the
$-x$ direction. The probability of the weak interaction of the
particle of the detector with a single neutrino is given by
\begin{equation}
w=|\langle\psi_{1}\psi_{2}|g|\psi_{\nu}\psi_{p}\rangle|^2
\label{eq:w}
\end{equation}
where $g$ is the weak coupling, $\psi_{p}$ is the state of the
particle of the detector, $\psi_{\nu}$ is the state of neutrino,
$\psi_{1}$, $\psi_{2}$
are the states of the particles arising in the interaction.
The probability given by eq.~(\ref{eq:w}) is too small to
register the event in a single interaction of the particle of the
detector with the neutrino. The probability of the event is a sum
of probabilities of a number of interactions.

To define the probability of interaction of the
particle of the detector with two neutrinos moving from the
opposite directions one should form the superposition of the
particle of the detector and neutrino states.
The state of the particle interacting with the neutrino moving from
the $x$ direction and the state of the particle interacting with the
neutrino moving from the $-x$ direction
are connected with the P-transformation
\begin{equation}
|\psi_{p}(-x)\rangle=P|\psi_{p}(x)\rangle.
\label{eq:P}
\end{equation}
Then the superposition of
the particle and neutrino states should be given by
\begin{eqnarray}
|\psi\rangle &=&
\displaystyle\frac{1}{\sqrt{2}}|\psi_{\nu}\rangle
[|\psi_{p}(x)\rangle+|\psi_{p}(-x)\rangle]\nonumber \\
\nonumber \\
&=& \displaystyle\frac{1}{\sqrt{2}}|\psi_{\nu}\rangle
[|\psi_{p}(x)\rangle+P|\psi_{p}(x)\rangle]\nonumber \\
\nonumber \\
&=& \displaystyle\frac{1}{\sqrt{2}}[|\psi_{\nu}\rangle+
P|\psi_{\nu}\rangle]|\psi_{p}(x)\rangle.
\label{eq:psi}
\end{eqnarray}
Since the P-transformation is forbidden for neutrino, one cannot
form the superposition given by eq.~(\ref{eq:psi}).
Hence only a half of neutrinos moving from the opposite directions
gives contribution into the superposition of the neutrino
states. Therefore the sum of probabilities of interaction of the
particle with neutrinos moving from the opposite directions is two
times smaller than that for the case when all neutrinos move
in one and the same direction
\begin{equation}
w_{\uparrow\downarrow}=0.5w_{\uparrow\uparrow}
\label{eq:w2}
\end{equation}

Let the detector measure the neutrino fluxes $F_i$ from $N$
sources placed in the plane $xy$. When not taking into account the
effect of the opposite neutrino directions given by
eq.~(\ref{eq:w2}) the total neutrino flux is
\begin{equation}
F=\sum\limits_{i=1}^{N}F_{i}.
\label{eq:F}
\end{equation}
To take into account the effect of the opposite neutrino
directions given by eq.~(\ref{eq:w2}) write
down the neutrino flux in the form
\begin{equation}
F_{i}=F_{ix}+F_{iy}
\label{eq:Fi}
\end{equation}
where
\begin{equation}
F_{ix}=F_{i}\cos^2\alpha
\label{eq:Fix}
\end{equation}
\begin{equation}
F_{iy}=F_{i}\sin^2\alpha
\label{eq:Fiy}
\end{equation}
where $\alpha$ is the angle between the direction to the source
and the axis $x$.
When taking into account the effect of the opposite neutrino
directions given by eq.~(\ref{eq:w2}) the total neutrino flux is
\begin{equation}
F^{*}=0.5\left(\sum\limits_{i=1}^{N}F_{i}+
\sum\limits_{i=1}^{N}\vec{F}_{ix}
+\sum\limits_{i=1}^{N}\vec{F}_{iy}\right)
\label{eq:Fv}
\end{equation}
where one should sum the fluxes $\vec{F}_{ix}$, $\vec{F}_{iy}$ as
vectors. One can see that the neutrino flux given by
eq.~(\ref{eq:Fv}) are within the range 0.5--1 of the neutrino
flux given by eq.~(\ref{eq:F}).

In the KamLAND experiment~\cite{Kam}, it is measured the flux of
$\bar{\nu}_e$'s from distant nuclear reactors.
The ratio of the number of observed events to the expected
number of events is reported~\cite{KC} to be
$0.611\pm 0.085 {\rm (stat)} \pm 0.041 {\rm (syst)} $.
Using the data on the location and thermal power of
reactors~\cite{insc}
one can estimate the neutrino flux from reactors which is
approximately proportional to the thermal power flux
$P_{th}/4\pi L^2$, where $P_{th}$ is the reactor thermal power,
$L$ is the distance from the detector to the reactor. Hence one
can estimate the effect of the opposite neutrino directions
through the ratio of the neutrino flux calculated with
eq.~(\ref{eq:Fv}) and that calculated with eq.~(\ref{eq:F}).
More than $99 \%$ of the neutrino flux measured in the KamLAND
experiment comes from Japanese and Korean nuclear reactors.
So in the first approximation one can calculate the neutrino flux
only from Japanese and Korean nuclear reactors.
Calculations for the conditions of the KamLAND experiment
yield the ratio of the neutrino flux taking into
account the effect of the opposite neutrino directions and that
without this effect, 0.555.
Thus the effect of the opposite neutrino directions may account
for the neutrino flux observed in the KamLAND experiment.
Note that the value 0.555 is obtained with the use of the
specification data on the reactor thermal power. To obtain the
real value one should use the operation data on the reactor
thermal power for the period of measurement.


\begin{thebibliography}{99}

\bibitem{Kam}
http://www.awa.tohoku.ac.jp/KamLAND/

\bibitem{KC}
K. Eguchi {\it et al.}, KamLAND Collaboration,
submitted to Phys. Rev. Lett., hep-ex/0212021.


\bibitem{insc}
http://www.insc.anl.gov/

\end{thebibliography}
\end{document}